\documentclass[11pt]{article}
\pdfoutput=1
\usepackage[a4paper,margin=1in]{geometry}
\usepackage{amsmath,amssymb,mathtools}
\usepackage{bm}
\usepackage{graphicx}
\usepackage{tabularx}
\usepackage{array}
\usepackage[numbers,sort&compress]{natbib}
\usepackage{hyperref}
\hypersetup{colorlinks=true,linkcolor=blue,citecolor=blue,urlcolor=blue}
\usepackage{xcolor}
\allowdisplaybreaks
\newcommand{\dd}{\mathrm{d}}

\newcommand{\cL}{\mathcal{L}}

\newcommand{\ren}{\mathrm{ren}}

\begin{document}
\begin{titlepage}

\hskip 1.5cm

\begin{center}
{\huge\bf{Covariant Holographic Entanglement Entropy Inversion to Reconstruct Bulk Geometry}}
\vskip 0.8cm
{\bf \large Ji-Seong Chae$^{\dag}$\footnote{ jiseongchae17@gmail.com}}
\vskip 0.75cm
{\em $^{\dag}$Department of Physics, Hanyang University, Seoul 04763, Korea}
\vspace{12pt}
\end{center}
\begin{abstract}
We derive an analytic inverse map from the renormalized covariant interval entropy to a symmetry-reduced stationary bulk metric.  The main result applies to stationary homogeneous asymptotically AdS$_3$ geometries with a radial lapse, an independent spatial warp factor, and a stationary shift.  On a smooth HRT branch, the interval entropy \(S_{\ren}(\Delta t,\Delta x)\) is an on-shell Hamilton--Jacobi functional.  Its endpoint derivatives determine two conserved charges, and the ratio \(\kappa=E/J\) organizes the extremal curves into one-parameter families with varying bulk turning points.  For each fixed \(\kappa\), the length and endpoint-separation data give three Volterra equations with the standard Abel square-root kernel. Their inverse kernels reconstruct the radial coordinate and the metric functions \(f,h,v\).  We also state the local consistency conditions and the agreement required when two local reconstructions of the same smooth HRT family cover an overlapping radial interval.  The construction is verified in pure AdS, rotating BTZ, a static warped geometry, and an Einstein--scalar black brane, and extended to higher-dimensional strips when the transverse area density is fixed by symmetry.
\end{abstract}

\vskip 1 cm
\vspace{24pt}
\end{titlepage}

\tableofcontents

\section{Introduction}

Holographic entanglement entropy relates the entropy of a boundary
region to the area of an associated bulk extremal surface.  For boundary
regions defined on a common time slice in a static or time-reflection
symmetric geometry, this relation is given by the Ryu--Takayanagi (RT)
prescription~\cite{Ryu:2006bv}.  Its covariant extension to general
Lorentzian geometries is the Hubeny--Rangamani--Takayanagi (HRT)
prescription~\cite{Hubeny:2007xt}. Within the context of  AdS/CFT~\cite{Maldacena:1997re,Gubser:1998bc,Witten:1998qj}, this relation provides a
direct example of how spacetime geometry is encoded in quantum entanglement ~\cite{Vidal:2007hda,Vidal:2008zz,Swingle:2009bg,VanRaamsdonk:2010pw,Maldacena:2013xja,Milsted:2018yur,Rangamani:2016dms,Lewkowycz:2013nqa,Faulkner:2013ica,Engelhardt:2014gca,Dong:2016fnf,Takayanagi:2017knl,Nguyen:2017yqw,Dutta:2019gen}.  Most applications of the RT/HRT formula
proceed in the forward direction: one specifies a bulk metric, determines
the corresponding extremal surfaces, and evaluates their areas to obtain the boundary entanglement entropy, as in early studies of
nontrivial entangling regions, strong subadditivity, and confining
backgrounds~\cite{Hirata:2006jx,Nishioka:2006gr,Headrick:2007km, Klebanov:2007ws}. This structure motivates the converse
question of whether the boundary entanglement data determine
the metric from which they arise.  The inverse problem is therefore to
identify which metric functions can be reconstructed, and under which
symmetry and regularity assumptions that reconstruction is unique.

Analytic progress on this problem has come from reducing it with symmetry. The main derivation concerns a stationary homogeneous asymptotically AdS$_3$ spacetime, written in a radial gauge as
\begin{equation}
  {
  \dd s^2
  ={1\over z^2}
  \left[
    G(z)\dd z^2
    -F(z)\dd t^2
    +H(z)\bigl(\dd x+V(z)\dd t\bigr)^2
  \right],
  }
  \label{eq:intro_general_metric}
\end{equation}
where the four coefficients depend on the radial coordinate alone.  The residual reparametrization $z\mapsto\tilde z(z)$ removes one of them; below it is fixed by $G=1/F$, and the three remaining functions are written $f=F$, $h=H$, and $v=V$.

The previous studies all sit inside the static subclass $V=0$, using the equal-time RT formula.  Hammersley extracted metric information from entanglement entropy for spherically symmetric geometries; the use of an areal radial coordinate fixes the angular warp factor, which corresponds to $H=1$ in Eq.~\eqref{eq:intro_general_metric}~\cite{Hammersley:2006cp,Hammersley:2007ab}.  Bilson treated the same class analytically~\cite{Bilson:2008ab}.  When the temporal and radial coefficients are related by $G=1/F$, one independent function remains and the metric can be reconstructed by boundary null geodesics.

Bilson's strip inversion assumes \(V=0\) and \(H=1\), while allowing \(F\) and \(G\) to remain independent~\cite{Bilson:2010ff}.  Because the
relevant RT surfaces lie on a constant-time slice, their induced metric contains \(G\) but not \(F\).  The strip entanglement entropy can therefore
reconstruct the radial coefficient \(G\), whereas the temporal coefficient
\(F\) remains undetermined and requires additional data sensitive to time
propagation.  A recent reconstruction extends this static program to thermodynamic and renormalization-group quantities, again using $V=0$, $H=1$, and $G=1/F$. In the thermal state, the ansatz contains one blackening function, while at zero temperature, boundary Lorentz invariance reduces the flow geometry to one warp factor~\cite{Huh:2026nvt}. After that metric function is reconstructed from entanglement entropy, the gravitational field equations can be used to infer the scalar potential, beta function, and $c$ function within the assumed Einstein--scalar model.

Other reconstruction programs use complementary observables and assumptions.  Light cone cuts and bulk cone singularities recover causal structure and, in favorable cases, the conformal metric~\cite{Hubeny:2006yu,Engelhardt:2016wgb,Engelhardt:2016xco}. Hamilton--Kabat--Lifschytz--Lowe (HKLL) reconstructs bulk fields once a background is specified~\cite{Hamilton:2006az}, while holographic renormalization fixes the expansion near the boundary from sources and one-point functions~\cite{deHaro:2000vlm}.  Kinematic space, differential entropy, and related integral geometric methods organize interval families~\cite{Balasubramanian:2013lsa,Czech:2014wka,Czech:2015qta,Myers:2014jia}; bit threads and entanglement wedge reconstruction provide global and operator algebraic descriptions of the same RT/HRT data~\cite{Freedman:2016zud,Dong:2016eik}.  Metric information has also been inferred from stress tensors, correlator singularities, Wilson loops, complexity, modular flow, perturbative area variations, numerical RT inversion, and inverse scattering~\cite{deHaro:2000vlm,Hubeny:2006yu,Engelhardt:2016wgb,Hashimoto:2020mrx,Hashimoto:2021kxx,Xu:2023,Kabat:2018smf,Roy:2018ehv,CaronHuot:2022,Bao:2019bib,Jokela:2025,Ahn:2025hre,Kim:2026irt,Fan:2025invscatt}.

Stationary states provide the first setting in which covariant interval data contain information that is absent from equal-time RT data.  In rotating BTZ, the left- and right-moving thermal scales encode the boundary angular momentum and produce a nonzero bulk shift~\cite{Banados:1992wn,Brown:1986nw,Maldacena:2001kr}.  More general rotating or hairy AdS black holes distinguish the horizon angular velocity   from the radial frame-dragging profile and the stationary limit surface~\cite{Berman:1999jq,Berman:2000rt,Hamaki:2025hair}.   The corresponding left--right asymmetry modifies the covariant interval entropy and provides boundary information sensitive to the stationary shift~\cite{Hubeny:2007xt,Asplund:2014coa}.

The present construction extends the inverse problem where $V\neq0$ and $H$ is arbitrary.  What makes this possible is that the covariant interval entropy depends on $\Delta t$ and $\Delta x$ separately, so Hamilton--Jacobi endpoint variation supplies two conserved charges rather than one.  Families with different charge ratios then separate all three coefficients of the stationary block, including the radial profile of frame dragging.

The renormalized holographic entropy \(S_{\ren}\) is treated, up to the
overall factor \(4G_N\), as an on-shell Hamilton--Jacobi functional.  Its
endpoint derivatives therefore determine the conserved energy and momentum
of the corresponding extremal surface,
\(E=-4G_N\partial_{\Delta t}S_{\ren}\) and
\(J=4G_N\partial_{\Delta x}S_{\ren}\).  The ratio
\(\kappa=E/J\) is held fixed while \(u_*=1/J\) moves the turning point.  At
fixed \(\kappa\), the entropy and the two endpoint separations reduce to
one-dimensional Volterra equations with a square-root kernel; throughout,
\textit{Abel inversion} denotes the inversion of equations of this type.

The construction then has three steps.  The entropy and its endpoint derivatives determine three Abel kernels; ratios of those kernels give the inverse metric block at each value of $\kappa$; and the families obtained at different $\kappa$ are compared at the same bulk radius.  They describe one stationary geometry only when they identify the same radial coordinate and depend on $\kappa$ in the way a single metric block requires.  Eqs.~\eqref{eq:F_inversion}--\eqref{eq:fvh_recon} then reconstruct $A,B,C$ and hence $f,h,v$.  The input is a smooth HRT branch on which the extremal problem reduces to one radial variational problem. Because the inverse map is local, the same branch may be reconstructed on overlapping interval domains; after the same radial gauge and asymptotic normalization are imposed, the resulting metric functions must agree on the radial range covered by both reconstructions.

The reconstructed functions also determine the null slopes tangent to a constant-$z$ stationary block, $\lambda_\pm(z)=-v(z)\pm\sqrt{f(z)/h(z)}$.  Their coalescence at a stationary horizon gives the direction of the horizon generator.

\begin{table}[t!]
\centering
\footnotesize
\renewcommand{\arraystretch}{1.18}
\caption{Scope of the inversion formula.  Here $N_{\rm unk}$ counts the gauge-inequivalent radial functions in the symmetry-reduced problem, and $N_{\rm data}$ counts the independent radial coefficient functions supplied by the chosen entropy family.}
\label{tab:scope_inversion}
\begin{tabularx}{\textwidth}{
  >{\raggedright\arraybackslash}p{0.20\textwidth}
  >{\raggedright\arraybackslash}X
  >{\centering\arraybackslash}p{0.08\textwidth}
  >{\centering\arraybackslash}p{0.09\textwidth}
  >{\raggedright\arraybackslash}p{0.22\textwidth}
}
\hline
Class & Representative reduced sector & $N_{\rm unk}$ & $N_{\rm data}$ & Result \\
\hline
static AdS$_3$ at equal boundary time
& $ds^2=z^{-2}[-f(z)dt^2+dz^2/f(z)+h(z)dx^2]$, $\Delta t=0$
& $2$
& $1$
& reconstructs one spatial combination; $f$ and $h$ are not separated
\\
static covariant AdS$_3$
& $ds^2=z^{-2}[-f(z)dt^2+dz^2/f(z)+h(z)dx^2]$
& $2$
& $2$
& reconstructs $f$ and $h$
\\
stationary homogeneous AdS$_3$
& $ds^2=z^{-2}[-f(z)dt^2+dz^2/f(z)+h(z)(dx+v(z)dt)^2]$
& $3$
& $3$
& reconstructs $f,h,v$
\\
isotropic equal-time strip in higher dimensions
& $ds^2_{\rm sp}=G(z)dz^2+H(z)d\vec x^{\,2}$
& $1$
& $1$
& reconstructs the reduced spatial metric after gauge fixing
\\
stationary isotropic covariant strip
& $W(z)^2[Gdz^2-N^2dt^2+H(dx+vdt)^2]$
& $3$
& $3$
& reconstructs the reduced stationary block when $W$ is fixed by isotropy
\\
anisotropic covariant strip, one orientation
& $W_i(z)^2[Gdz^2-N^2dt^2+H_i(dx_i+v_i dt)^2]$
& $>3$
& $3$
& reconstructs only the $W_i^2$-dressed stationary block
\\
time-dependent or inhomogeneous geometry
& $ds^2=g_{\mu\nu}(t,z,x^i)dx^\mu dx^\nu$
& not radial
& --
& outside the present one-dimensional inversion
\\
\hline
\end{tabularx}
\end{table}

Table~\ref{tab:scope_inversion} summarizes the sectors supported by the derivation. The relevant count is the number $N_{\rm unk}$ of gauge-inequivalent radial functions that remain after the symmetry reduction, compared with the number $N_{\rm data}$ of independent coefficient functions carried by the chosen HRT family.  Covariant intervals close the static two-function problem and the stationary three-function problem in AdS$_3$.  Higher-dimensional isotropic strips also close after the transverse area density is fixed by symmetry, whereas one anisotropic orientation determines only their products and cannot separate the transverse area density from the radial metric functions.  The examples of Section~\ref{sec:examples} test the radial normalization, a nonzero stationary shift, an independent spatial warp factor $h(z)\neq1$, and a matter-supported geometry.

The paper is organized as follows.  Section~\ref{sec:review}  reviews the RT/HRT prescription and the static geodesic and strip inversions that motivate the covariant construction.  Section~\ref{sec:inversion} derives the covariant inverse map, states the local consistency and overlap conditions, and discusses the higher-dimensional strip reduction.  Section~\ref{sec:examples} presents the explicit checks.  Section~\ref{sec:discussion} summarizes the result, its limitations, and possible extensions.

\section{Holographic entanglement entropy and static inversions}\label{sec:review}

\subsection{From Ryu--Takayanagi to Hubeny--Rangamani--Takayanagi}\label{subsec:rt_hrt_review}

For a static asymptotically anti--de Sitter (AdS) bulk spacetime, the Ryu--Takayanagi (RT) prescription assigns to a spatial boundary region $A$ the entropy
\begin{equation}
  S(A)={\mathrm{Area}(\gamma_A)\over 4G_N},
  \label{eq:RT_formula}
\end{equation}
where $\gamma_A$ is the bulk codimension-two minimal surface anchored on $\partial A$ and homologous to $A$~\cite{Ryu:2006bv}.  In AdS$_3$ the surface is a spacelike geodesic, so the area is a length.  We write
\begin{equation}
  L_{\ren}(A)\equiv4G_N S_{\ren}(A),
  \label{eq:Sren_def}
\end{equation}
for the renormalized length after subtraction of the universal divergence near the boundary.

For a spacelike codimension-two surface \(\mathcal X_A\) in a Lorentzian bulk, choose null normals \(k_+^\mu\) and \(k_-^\mu\).  Their expansions
\begin{equation}
  \theta_{(\pm)}
  =
  \frac{1}{\sqrt{\gamma}}\mathcal L_{k_\pm}\sqrt{\gamma}
\end{equation}
measure the first-order change of the area density under normal displacement.  The Hubeny--Rangamani--Takayanagi (HRT) surface is the Lorentzian extremal surface satisfying
\begin{equation}
  \theta_{(+)}\big|_{\mathcal X_A}=0,
  \qquad
  \theta_{(-)}\big|_{\mathcal X_A}=0,
\end{equation}
with \(\partial\mathcal X_A=\partial A\).  The area is stationary under both independent normal deformations, and the physical saddle is the extremal surface of least area among the competing homologous extrema.

In Lorentzian signature the local condition is extremality under both normal directions.  For a static geometry \(g_{tx}=0\) invariant under time reflection, the problem can be represented as a minimization on the invariant spatial slice.  A stationary geometry with \(g_{tx}\neq0\) has no such invariant slice in general, so its extremal curve must be obtained from the covariant HRT variational problem.

\subsection{Static geodesic and strip inversions}\label{subsec:bilson_review}

We review the two static reference inversions.  The first combines null and spacelike geodesic data to separate two radial functions. The second uses HEE for equal-time strip regions and shows explicitly which metric coefficient such data can probe.

\medskip
\paragraph{Geodesic inversion.}
Bilson's first construction uses endpoint data for null geodesics that return to the boundary and, in AdS$_3$, spacelike geodesic lengths~\cite{Bilson:2008ab}.  The one-function example starts from
\begin{equation}
  \dd s^2=-F(r)\dd t^2+{\dd r^2\over F(r)}+r^2\dd\Omega_n^2 .
  \label{eq:bilson_one_function_metric}
\end{equation}
For a null geodesic, spherical symmetry allows the motion to be restricted to a great circle with angular coordinate {$\phi$}.  The Killing fields $\partial_t$ and $\partial_\phi$ give
\begin{equation}
  E\equiv-p_t=F(r)\dot t,
  \qquad
  J\equiv p_\phi=r^2\dot\phi ,
\end{equation}
where the dot denotes differentiation with respect to an affine parameter.  The inverse impact parameter is
\begin{equation}
  \alpha\equiv{E\over J}.
\end{equation}

The null condition gives
\begin{align}
  \dot r^2
  =E^2-{F(r)J^2\over r^2}
  =J^2\left(\alpha^2-{F(r)\over r^2}\right).
\end{align}
The turning point $r_{\min}$ is fixed by
\begin{equation}
  \alpha^2={F(r_{\min})\over r_{\min}^2}.
\end{equation}
The corresponding boundary separations are
\begin{equation}
  \Delta t(\alpha)=2\int_{r_{\min}(\alpha)}^{\infty}
  {\alpha\,\dd r\over F(r)\sqrt{\alpha^2-F(r)/r^2}},
  \label{eq:bilson_delta_t}
\end{equation}
\begin{equation}
  \Delta\phi(\alpha)=2\int_{r_{\min}(\alpha)}^{\infty}
  {\dd r\over r^2\sqrt{\alpha^2-F(r)/r^2}}.
  \label{eq:bilson_delta_phi}
\end{equation}
  Their endpoint variation satisfies
  \begin{equation}
  {\dd\over\dd\alpha}\bigl[\alpha\Delta t(\alpha)\bigr]
  ={\dd\over\dd\alpha}\Delta\phi(\alpha).
  \label{eq:bilson_dt_dphi_relation}
\end{equation}
Thus, one boundary function determines the single radial function.  Defining
\begin{equation}
  V(r)\equiv{F(r)\over r^2},
  \label{eq:bilson_potential}
\end{equation}
and changing variables from $r$ to $V$,   the angular separation becomes
\begin{equation}
  \Delta\phi(\sqrt{V_{\max}})
  =\int_{1}^{V_{\max}}{p(V)\,\dd V\over\sqrt{V_{\max}-V}},
  \qquad
  p(V)\equiv-2{\dd r\over\dd V}{1\over r(V)^2}.
  \label{eq:bilson_abel_null}
\end{equation}
Applying the Abel inversion determines $p(V)$, then $r(V)$, and finally {$F(r)=r^2V(r)$.  The reconstruction is local to a monotone turning-point branch.

For a static metric with two independent radial functions, write
\begin{equation}
  \dd s^2=-F(r)\dd t^2+{\dd r^2\over G(r)}+r^2\dd\Omega_n^2 .
  \label{eq:bilson_two_function_metric}
\end{equation}
With $V(r)\equiv F(r)/r^2$, the null radial equation is
\begin{equation}
  \dot r^2
  ={G(r)J^2\over r^2V(r)}\left(\alpha^2-V(r)\right),
\end{equation}
and the angular separation is
\begin{equation}
  \Delta\phi(\alpha)
  =2\int_{r_{\min}(\alpha)}^{\rho_0}
  {\sqrt{V(r)}\,\dd r\over r\sqrt{G(r)}\sqrt{\alpha^2-V(r)}} .
  \label{eq:bilson_two_function_null_angle}
\end{equation}
The null data determine only the radial combination
\begin{equation}
  \mathcal I_1(V(r))
  \equiv-{1\over rV'(r)}\sqrt{{V(r)\over G(r)}} ,
  \label{eq:bilson_I1_radial_combination}
\end{equation}
whose inverse Abel representation is
\begin{equation}
  \mathcal I_1(y)
  =-{1\over\pi}{\dd\over\dd y}
  \int_1^{\sqrt y}{\alpha\Delta\phi(\alpha)\,\dd\alpha\over\sqrt{y-\alpha^2}}.
  \label{eq:bilson_I1_def}
\end{equation}
The second function is supplied by equal-time spacelike geodesic lengths.  For unit-speed normalization, the turning point is $r=J$ and
\begin{equation}
  L(J)=2\int_J^{\rho_0}{r\,\dd r\over\sqrt{G(r)(r^2-J^2)}} .
  \label{eq:bilson_length_h}
\end{equation}
The Abel inversion gives
\begin{equation}
  {1\over\sqrt{G(r)}}
  ={1\over r\pi}{\dd\over\dd r}
  \int_{\rho_0}^{r}{J L(J)\,\dd J\over\sqrt{J^2-r^2}} .
  \label{eq:bilson_G_inversion}
\end{equation}
Once $G(r)$ is known,   Eq.~\eqref{eq:bilson_I1_radial_combination} becomes the first-order equation
\begin{equation}
  V'(r)+{1\over r\,\mathcal I_1(V(r))}\sqrt{{V(r)\over G(r)}}=0,
  \label{eq:bilson_F_after_G}
\end{equation}
 with the asymptotically AdS boundary condition.  The spacelike lengths therefore determine $G(r)$, while the null endpoint data then determine $V(r)$ and $F(r)=r^2V(r)$.

\medskip
\phantomsection\label{subsec:obstruction}
\paragraph{Strip inversion and the equal-time obstruction.}
Bilson's second construction considers static planar metrics
~\cite{Bilson:2010ff}
\begin{equation}
  \dd s^2={R^2\over z^2}
  \left[-N(z)^2\dd t^2+G(z)^2\dd z^2+\sum_{i=1}^d\dd x_i^2\right].
  \label{eq:bilson_part2_metric}
\end{equation}
For a strip
\begin{equation}
  A_S=\{x_1=x\in[-\ell/2,\ell/2],\;x_2,\ldots,x_d\in\mathbb R\},
  \label{eq:bilson_strip_region}
\end{equation}
the embedding $z=z(x)$ has area
\begin{equation}
  A_\gamma(\ell)=R^dL^{d-1}\int_{-\ell/2}^{\ell/2}
  {\sqrt{1+G(z)^2z'(x)^2}\over z^d}\,\dd x.
  \label{eq:bilson_strip_area}
\end{equation}
Since the area density has no explicit $x$-dependence, the Beltrami identity gives
\begin{equation}
  \mathcal A-z'{\partial\mathcal A\over\partial z'}
  ={1\over z^d\sqrt{1+G(z)^2z'^2}}
  ={1\over z_*^d},
  \qquad
  \mathcal A(z,z')\equiv{\sqrt{1+G(z)^2z'^2}\over z^d},
  \label{bidentity}
\end{equation}
where $z=z_*$ and $z'=0$ at the turning point.  Hence
\begin{equation}
  {\dd z\over\dd x}
  ={\sqrt{z_*^{2d}-z^{2d}}\over z^dG(z)}.
  \label{eq:bilson_strip_profile}
\end{equation}
The strip width and area are
\begin{equation}
  \ell(z_*)=2\int_0^{z_*}{z^dG(z)\,\dd z\over\sqrt{z_*^{2d}-z^{2d}}},
  \label{eq:bilson_strip_width}
\end{equation}
\begin{equation}
  A_\gamma(z_*)=2R^dL^{d-1}
  \int_a^{z_*}{z_*^dG(z)\,\dd z\over z^d\sqrt{z_*^{2d}-z^{2d}}}.
  \label{eq:bilson_strip_area_turning}
\end{equation}
The boundary input $S_A(l\ell)$ is parametrized by the bulk turning point $z_*$. To convert the area as inverse data, one therefore first needs a metric-independent map between $\ell$ and $z_*$. The on-shell variation with respect to the strip width gives
\begin{equation}
  {\dd S_A\over\dd\ell}={R^dL^{d-1}\over2G_N^{(d+2)}z_*^d}.
  \label{eq:bilson_dSdl_map}
\end{equation}
Hence, $S_Z(\ell)$ determines $z_*(\ell)$, and therefore $A_\gamma(z_*)$. Applying the Abel inversion gives
\begin{equation}
  G(z)={d z^d\over\pi R^dL^{d-1}}{\dd\over\dd z}
  \int_a^z{A_\gamma(z_*) z_*^{d-1}\,\dd z_*\over\sqrt{z^{2d}-z_*^{2d}}}.
  \label{eq:bilson_part2_G_inversion}
\end{equation}
 Thus, equal-time strip data determine the spatial radial coefficient $G(z)$ but do not probe the lapse $N(z)$.  Related static reconstructions, including applications to black holes and matter fields after a gravitational ansatz is imposed, were developed in Ref.~\cite{Huh:2026nvt}.

\subsection{From static probes to covariant HRT}\label{subsec:covariant_review}

A spacelike boundary interval in a Lorentzian CFT has independent separations \((\Delta t,\Delta x)\).  In a stationary homogeneous three-dimensional bulk, endpoint variation of the renormalized HRT length gives two conserved momenta,
\begin{equation}
  E\equiv-\partial_{\Delta t}L_{\ren},
  \qquad
  J\equiv\partial_{\Delta x}L_{\ren},
  \qquad
  \kappa\equiv{E\over J},
  \qquad
  u_*\equiv{1\over J}.
  \label{eq:review_HJ_pair}
\end{equation}
Here \(\kappa=E/J\) fixes the ratio of the two charges, and the inverse spatial momentum \(u_*=1/J\) varies the turning point while that ratio is held fixed.  Thus, for each value of \(\kappa\), the HRT data contain a family of curves that share the same charge ratio and reach different radial depths.  Figure~\ref{fig:charge_family_organization} summarizes this organization.  The intervals at equal boundary time form the \(\kappa=0\) slice; intervals with nonzero \(\kappa\) provide the additional covariant information used in the reconstruction.

\begin{figure}[t!]
  \centering
  \includegraphics[width=\textwidth]{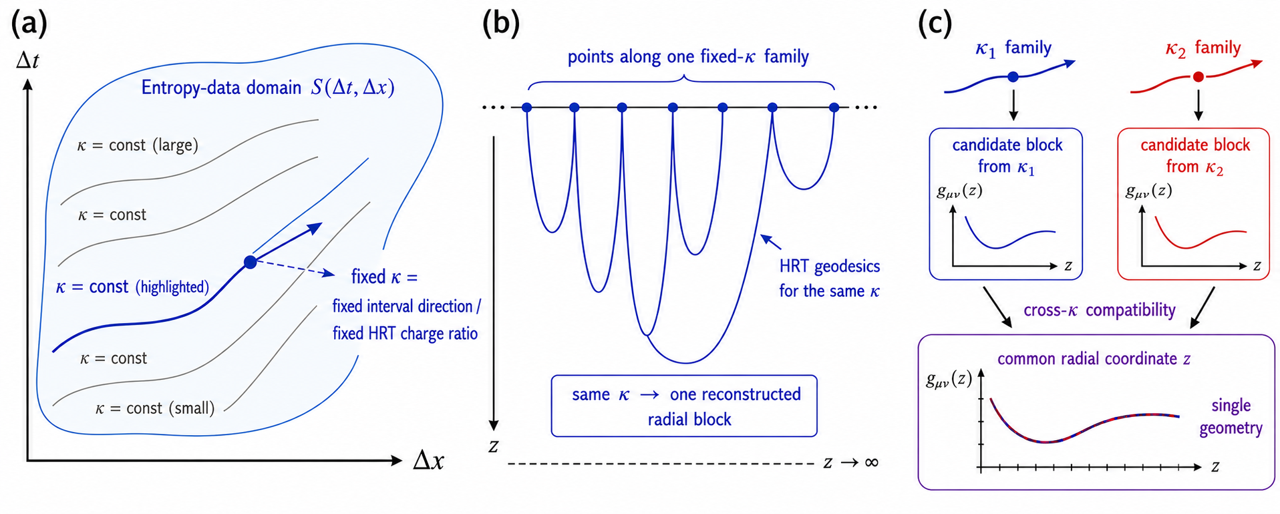}
  \caption{Schematic diagram of the organization of the covariant inverse problem.  Panel (a) groups the entropy data \(S_{\ren}(\Delta t,\Delta x)\) according to the Hamilton--Jacobi ratio \(\kappa=E/J\), with \(E=-\partial_{\Delta t}L_{\ren}\) and \(J=\partial_{\Delta x}L_{\ren}\).  Panel (b) shows that varying \(u_*=1/J\) while \(\kappa\) is held fixed changes the turning point and supplies the data for the Abel inversion.  Panel (c) shows the consistency requirement: all values of \(\kappa\) must lead to the same radial coordinate \(z\) and the same metric functions.}
  \label{fig:charge_family_organization}
\end{figure}
The construction is local on a smooth HRT family.  We restrict to an open interval domain on which the selected extremal length $L(\Delta t,\Delta x)$ is three times continuously differentiable and the Hamilton--Jacobi variables provide local coordinates:
\begin{equation}
  J\neq0,
  \qquad
  \det {\partial(\kappa,u_*)\over \partial(\Delta t,\Delta x)}\neq0 .
  \label{eq:HJ_non_degenerate}
\end{equation}
The coordinate description fails when $J=0$ or when the Jacobian vanishes, as may occur at a caustic or a degenerate turning point.  If several extremal surfaces compete, the physical entropy selects the one with the least area.  The present inversion applies where the selected extremum varies smoothly and does not determine the lengths of the other extrema across an HRT } phase transition.

Section~\ref{sec:inversion} converts the data in Eq.~\eqref{eq:review_HJ_pair} into three Abel kernels.  These kernels reconstruct the three components of the inverse metric block when the families at different values of \(\kappa\) identify one radial coordinate and obey the required dependence on the charge ratio.  The static inversion with one unknown function is recovered as a restricted limit.

\section{Inversion formula from covariant holographic entanglement entropy}\label{sec:inversion}

This section derives the inverse map for the stationary homogeneous sector stated in the Introduction.  The covariant interval entropy supplies two Hamilton--Jacobi charges.  Their ratio \(\kappa=E/J\) is held fixed, while \(u_*=1/J\) varies the radial turning point.  The derivation first applies the Abel transform with \(\kappa\) held fixed and then imposes the consistency requirements needed for all values of \(\kappa\) to describe one radial metric block.

\subsection{Stationary metric class}

\phantomsection\label{sec:metricclass}

Consider the stationary homogeneous metric in a radial slicing
\begin{equation}
  \dd s^2
  ={1\over z^2}\left[G(z)\dd z^2+\gamma_{ab}(z)\dd x^a\dd x^b\right],
  \qquad x^a\equiv(t,x).
  \label{eq:general_3d_radial}
\end{equation}
This form does not completely fix the radial coordinate, since it
is preserved by residual reparametrizations \(z\mapsto\tilde z(z)\)
accompanied by corresponding redefinitions of \(G\) and
\(\gamma_{ab}\).  This freedom will be fixed below before the radial
functions are reconstructed.

Every Lorentzian two-dimensional block with $\gamma_{xx}>0$ admits the decomposition
\begin{equation}
  \gamma_{ab}\dd x^a\dd x^b
  =-N(z)^2\dd t^2+H(z)\left(\dd x+V(z)\dd t\right)^2.
  \label{eq:square_completion_general}
\end{equation}
This is the Arnowitt--Deser--Misner parametrization of the $(t,x)$ block within the stationary homogeneous sector.

For the explicit inversion, we fix the residual radial freedom by
imposing \(G(z)N(z)^2=1\) and write
\begin{equation}
  N(z)^2\equiv f(z),
  \qquad
  H(z)\equiv h(z),
  \qquad
  V(z)\equiv v(z),
  \qquad
  G(z)={1\over f(z)}.
  \label{eq:schwarzschild_gauge_choice}
\end{equation}
In this gauge the metric is
\begin{equation}
  \dd s^2
  ={1\over z^2}\left[-f(z)\dd t^2+{\dd z^2\over f(z)}+h(z)\left(\dd x+v(z)\dd t\right)^2\right].
  \label{eq:main_metric_convention}
\end{equation}
This gauge is convenient for the BTZ and black brane examples.

With the convention in Eq.~\eqref{eq:main_metric_convention}, define the components of the inverse metric block
\begin{equation}
  A(z)\equiv{1\over f(z)},
  \qquad
  B(z)\equiv{v(z)\over f(z)},
  \qquad
  C(z)\equiv{1\over h(z)}-{v(z)^2\over f(z)}.
  \label{eq:ABC_def}
\end{equation}
These functions are the components of the inverse of the two-dimensional metric $\gamma_{ab}$ of the $(t,x)$ coordinate and are the combinations that enter the geodesic Hamiltonian:
\begin{equation}
  \gamma^{tt}=-A,
  \qquad
  \gamma^{tx}=B,
  \qquad
  \gamma^{xx}=C.
  \label{eq:inverse_block}
\end{equation}
The metric functions are recovered algebraically from $A,B,C$:
\begin{equation}
  f={1\over A},
  \qquad
  v={B\over A},
  \qquad
  h={1\over C+B^2/A}.
  \label{eq:metric_from_ABC}
\end{equation}
The inverse problem is therefore the reconstruction of $A,B,C$ in the chosen radial gauge.  We use the sign convention $\dd x+v\dd t$ throughout.

\subsection{Characteristic Abel inversion and metric reconstruction}

\phantomsection\label{sec:abel}

Let $s$ denote proper length along the spacelike HRT geodesic, so that
\begin{equation}
  g_{\mu\nu}\dot X^\mu\dot X^\nu=\frac{1}{z^2}\left[\frac{\dot z^2}{f}-f\dot t^2+h(\dot x+v \dot t)^2\right]=1,
    \label{eq:proper_norm}
\end{equation}
where a dot denotes $\dd/\dd s$ and its canonical momentum is $p_\mu\equiv g_{\mu\nu} \dot x^\nu$.
The Killing fields $\partial_t$ and $\partial_x$ give the conserved energy $E=-p_t$ and spatial momentum $J=p_x$ as
\begin{equation}
  E\equiv-(g_{tt} \dot t+g_{tx} \dot x)={ (f-hv^2)\dot t-hv\dot x\over z^2},
  \qquad
  J\equiv g_{xt} \dot t+g_{xx} \dot x={ hv\dot t+h\dot x\over z^2}.
  \label{eq:EJ_def}
\end{equation}
Solving these relations for the velocities gives
\begin{equation}
  \dot t=z^2{E+vJ\over f},
  \qquad
  \dot x=z^2{J\over h}-z^2v{E+vJ\over f}.
  \label{eq:tdot_xdot}
\end{equation}
Substituting the conserved charges into Eq.~\eqref{eq:proper_norm} gives
\begin{equation}
  \dot z^2
  =z^2 f(z)\left[1-z^2\left({J^2\over h(z)}-{(E+v(z)J)^2\over f(z)}\right)\right].
  \label{eq:radial_eq_pre}
\end{equation}
Introduce the charge ratio $\kappa={E\over J}$ and its radial symbol,
\begin{equation}
  \Psi_\kappa(z)\equiv C(z)-2\kappa B(z)-\kappa^2A(z)=\frac{1}{h}-\frac{v^2}{f}-2\kappa \frac{v}{f}-\kappa^2\frac{1}{f}.
  \label{eq:Psi_def}
\end{equation}
Using $E=\kappa J$,
\begin{equation}
  {J^2\over h}-{(E+vJ)^2\over f}
  =J^2\Psi_\kappa(z),
  \label{eq:Psi_identity}
\end{equation}
and the radial equation takes the form
\begin{equation}
  \dot z^2=z^2f(z)\left[1-J^2z^2\Psi_\kappa(z)\right].
  \label{eq:radial_eq}
\end{equation}

At a regular radial turning point $z=z_*$,
\begin{equation}
  \dot z(z_*)=0.
  \label{eq:turning_condition}
\end{equation}
If the turning point lies outside a zero of $f$ on the selected branch, Eq.~\eqref{eq:radial_eq} gives
\begin{equation}
  1-J^2z_*^2\Psi_\kappa(z_*)=0.
  \label{eq:turning_point_equation}
\end{equation}
Define the characteristic radial variable
\begin{equation}
  u_\kappa(z)\equiv z\sqrt{\Psi_\kappa(z)}.
  \label{eq:u_def}
\end{equation}
The condition at the turning point is then
\begin{equation}
  J^2u_\kappa(z_*)^2=1.
  \label{eq:J_u_relation}
\end{equation}
For the orientation $J>0$,
\begin{equation}
  u_*=u_\kappa(z_*)={1\over J}.
  \label{eq:u_star_turning}
\end{equation}
Hence, the characteristic radial variable in Eq.~\eqref{eq:u_def} is the geometric representation of the inverse spatial momenta in Eq.~\eqref{eq:review_HJ_pair}.  For $J<0$ one replaces $1/J$ by $1/|J|$ and keeps the corresponding orientation.  We use $J>0$ below.

Eqs.~\eqref{eq:tdot_xdot} and \eqref{eq:radial_eq} give
\begin{equation}
  {\dd t\over\dd z}
  ={Jz\sqrt{A(z)}\left(\kappa A(z)+B(z)\right)
  \over \sqrt{1-J^2z^2\Psi_\kappa(z)}},
  \label{eq:dtdz}
\end{equation}
and
\begin{equation}
  {\dd x\over\dd z}
  ={Jz\sqrt{A(z)}\left(C(z)-\kappa B(z)\right)
  \over \sqrt{1-J^2z^2\Psi_\kappa(z)}}.
  \label{eq:dxdz}
\end{equation}
For a geodesic symmetric about its turning point, the endpoint separations are twice the integrals over one radial branch:
\begin{align}
  {\Delta t_\kappa\over2}
  &=J\int_0^{z_*}
  {z\sqrt{A}(\kappa A+B)\over\sqrt{1-J^2u_\kappa(z)^2}}\,\dd z,
  \label{eq:Delta_t_z}\\
  {\Delta x_\kappa\over2}
  &=J\int_0^{z_*}
  {z\sqrt{A}(C-\kappa B)\over\sqrt{1-J^2u_\kappa(z)^2}}\,\dd z.
  \label{eq:Delta_x_z}
\end{align}
Similarly, since \(\dot z=dz/ds\), changing variables in the integral of proper length \(L=\int ds\) gives \(ds=dz/|\dot z|\) on each radial branch.  The regulated geodesic length is therefore
\begin{align}
  L_\kappa(\epsilon)
  =
  2\int_\epsilon^{z_*}{dz\over |\dot z|}
  =
  2\int_{\epsilon}^{z_*}
  {\sqrt{A(z)}\over z\sqrt{1-J^2u_\kappa(z)^2}}\,dz .
\end{align}
The asymptotically AdS endpoint divergence is removed by defining
\begin{align}
  L_{\kappa,\rm ren}
  =
  \lim_{\epsilon\to0}
  \left[
  L_\kappa(\epsilon)+L_{\rm ct}(\epsilon)
  \right],
\end{align}
where $L_{\rm ct}$ is the standard local counterterm.  It removes the endpoint term that depends on the cutoff and fixes the finite normalization of the entropy input so that it does not modify the Abel transform.

Assume that the characteristic variable is monotone on the selected branch,
\begin{equation}
  u_\kappa'(z)\neq0,
  \label{eq:u_monotone}
\end{equation}
so that $z=z_\kappa(u)$ is well defined locally at fixed $\kappa$.  Since $J=1/u_*$,
\begin{equation}
  \sqrt{1-J^2u^2}={\sqrt{u_*^2-u^2}\over u_*}.
  \label{eq:denom_abel}
\end{equation}
Changing variables from $z$ to $u$ gives the length equation
\begin{equation}
  L_\kappa(u_*)
  =2u_*\int_0^{u_*}{F_\kappa(u)\over\sqrt{u_*^2-u^2}}\,\dd u,
  \label{eq:L_abel}
\end{equation}
where
\begin{equation}
  F_\kappa(u)\equiv\left.{\sqrt{A(z)}\over z\,u_\kappa'(z)}\right|_{z=z_\kappa(u)}.
  \label{eq:F_def}
\end{equation}
The time and spatial endpoint equations in Eqs.~\eqref{eq:Delta_t_z} and \eqref{eq:Delta_x_z} become
\begin{align}
  {\Delta t_\kappa(u_*)\over2}
  &=\int_0^{u_*}{T_\kappa(u)\over\sqrt{u_*^2-u^2}}\,\dd u,
  \label{eq:Dt_abel}\\
  {\Delta x_\kappa(u_*)\over2}
  &=\int_0^{u_*}{X_\kappa(u)\over\sqrt{u_*^2-u^2}}\,\dd u,
  \label{eq:Dx_abel}
\end{align}
with
\begin{equation}
  T_\kappa(u)\equiv\left.{z\sqrt{A(z)}\left(\kappa A(z)+B(z)\right)
  \over u_\kappa'(z)}\right|_{z=z_\kappa(u)},
  \label{eq:T_def}
\end{equation}
and
\begin{equation}
  X_\kappa(u)\equiv\left.{z\sqrt{A(z)}\left(C(z)-\kappa B(z)\right)
  \over u_\kappa'(z)}\right|_{z=z_\kappa(u)}.
  \label{eq:X_def}
\end{equation}
The same geodesic Hamiltonian, the turning point relation $J=1/u_*$, and the change of variables $z\mapsto u_\kappa(z)$ therefore determine all three kernels.

The corresponding inverse Abel formulae are
\begin{equation}
  F_\kappa(u)
  ={1\over\pi}{\dd\over\dd u}
  \int_0^u {2y\,\cL_\kappa(y)\over\sqrt{u^2-y^2}}\,\dd y,
  \qquad
  \cL_\kappa(u)\equiv{L_\kappa(u)\over2u},
  \label{eq:F_inversion}
\end{equation}
\begin{equation}
  T_\kappa(u)
  ={1\over\pi}{\dd\over\dd u}
  \int_0^u {y\,\Delta t_\kappa(y)\over\sqrt{u^2-y^2}}\,\dd y,
  \label{eq:T_inversion}
\end{equation}
\begin{equation}
  X_\kappa(u)
  ={1\over\pi}{\dd\over\dd u}
  \int_0^u {y\,\Delta x_\kappa(y)\over\sqrt{u^2-y^2}}\,\dd y.
  \label{eq:X_inversion}
\end{equation}

 To derive the metric block $A,B,C$ explicitly, it is useful to define the following ratios:
\begin{equation}
  R_t(\kappa,u)\equiv {T_\kappa(u)\over F_\kappa(u)},
  \qquad
  R_x(\kappa,u) \equiv {X_\kappa(u)\over F_\kappa(u)}.
  \label{eq:R_def}
\end{equation}
Then Eqs.~\eqref{eq:F_def}--\eqref{eq:X_def} give the pointwise identities
\begin{equation}
  R_t=z^2(\kappa A+B),
  \label{eq:Rt_identity}
\end{equation}
\begin{equation}
  R_x=z^2(C-\kappa B),
  \label{eq:Rx_identity}
\end{equation}
and therefore
\begin{equation}
  R_x-\kappa R_t=z^2\Psi_\kappa(z)=u^2.
  \label{eq:consistency_identity}
\end{equation}

The Abel kernels are functions of $(\kappa,u)$, whereas $A$, $B$, and
$C$ are functions of the bulk radius $z$.  To compare neighboring
$\kappa$-families at the same bulk point, the variation with respect to
$\kappa$ must therefore be taken at fixed $z$. The derivative with respect to \(\kappa\) at fixed $z$ follows from
\begin{equation}
  u^2=z^2\Psi_\kappa(z),
  \label{eq:u2_again}
\end{equation}
we find
\begin{equation}
  \left(\partial_\kappa u^2\right)_z
  =z^2\partial_\kappa\Psi_\kappa
  =-2z^2(\kappa A+B)
  =-2R_t.
  \label{eq:du2dk}
\end{equation}
Hence
\begin{equation}
  \left(\partial_\kappa u\right)_z=-{R_t\over u}.
  \label{eq:dudk_fixed_z}
\end{equation}
The chain rule then gives the derivative with respect to the charge ratio at fixed $z$
\begin{equation}
  D_\kappa\equiv\left(\partial_\kappa\right)_z
  =\partial_\kappa-{R_t\over u}\partial_u.
  \label{eq:Dkappa_def}
\end{equation}
Applied to Eq.~\eqref{eq:Rt_identity}, it gives
\begin{equation}
  D_\kappa R_t=z^2A.
  \label{eq:DkRt}
\end{equation}
Applied to Eq.~\eqref{eq:Rx_identity}, it gives
\begin{equation}
  D_\kappa R_x=-z^2B.
  \label{eq:DkRx}
\end{equation}
These identities separate the three coefficients of the stationary inverse block by comparing nearby values of \(\kappa\) at the same radial position.

In the gauge of Eq.~\eqref{eq:main_metric_convention}, $G=A$, so Eq.~\eqref{eq:F_def} becomes
\begin{equation}
  F_\kappa={\sqrt A\over z\,u_\kappa'(z)}.
  \label{eq:F_GA}
\end{equation}
Eq.~\eqref{eq:DkRt} gives
\begin{equation}
  \sqrt A={\sqrt{D_\kappa R_t}\over z}.
  \label{eq:sqrtA_DkRt}
\end{equation}
Using $\partial_u z=1/u_\kappa'(z)$ yields
\begin{equation}
  \partial_u z={F_\kappa z^2\over\sqrt{D_\kappa R_t}},
  \label{eq:z_ode}
\end{equation}
or, equivalently,
\begin{equation}
  \partial_u\left({1\over z}\right)
  =-{F_\kappa(u)\over\sqrt{D_\kappa R_t(\kappa,u)}}.
  \label{eq:one_over_z_ode}
\end{equation}
The asymptotically AdS boundary condition fixes the radial integration constant.  In a boundary frame with vanishing asymptotic shift, the gauge of Eq.~\eqref{eq:main_metric_convention} requires
\begin{align}
  f(z)\to1,\qquad h(z)\to1,\qquad v(z)\to0
  \qquad (z\to0).
\end{align}
Equivalently,
\begin{align}
  A\to1,\qquad B\to0,\qquad C\to1 .
\end{align}
The definition of $u$ then gives
\begin{align}
  u^2\sim z^2(1-\kappa^2),
  \qquad
  u\sim z\sqrt{1-\kappa^2}.
\end{align}
Thus
\begin{align}
  {1\over z}\sim{\sqrt{1-\kappa^2}\over u}
  \qquad (u\to0).
\end{align}
The same boundary normalization implies
\begin{align}
  {F_\kappa(u)\over\sqrt{D_\kappa R_t(\kappa,u)}}
  \sim
  {\sqrt{1-\kappa^2}\over u^2}
\end{align}
near $u=0$.  Subtracting this universal term gives the finite integral form
\[
  {1\over z(\kappa,u)}
  =
  {\sqrt{1-\kappa^2}\over u}
  -
  \int_0^u
  \left[
  {F_\kappa(s)\over\sqrt{D_\kappa R_t(\kappa,s)}}
  -
  {\sqrt{1-\kappa^2}\over s^2}
  \right]ds .
\]
The bracketed integrand is finite for the standard asymptotically AdS expansion.

Once $z(\kappa,u)$ has been fixed, Eqs.~\eqref{eq:DkRt}, \eqref{eq:Rt_identity}, and \eqref{eq:Rx_identity} reconstruct
\begin{equation}
  A={D_\kappa R_t\over z^2},
  \label{eq:A_recon}
\end{equation}
\begin{equation}
  B={R_t-\kappa D_\kappa R_t\over z^2}
  =-{D_\kappa R_x\over z^2},
  \label{eq:B_recon}
\end{equation}
The equality of the two expressions is equivalent to
\begin{equation}
  D_\kappa R_x+R_t-\kappa D_\kappa R_t=0,
  \label{eq:R_derivative_relation}
\end{equation}
which follows from Eqs.~\eqref{eq:Rt_identity}, \eqref{eq:Rx_identity}, \eqref{eq:DkRt}, and \eqref{eq:DkRx}.
The remaining diagonal component follows by solving
Eq.~\eqref{eq:Rx_identity}, $R_x=z^2(C-\kappa B)$, for $C$ and
substituting Eq.~\eqref{eq:B_recon}:
\begin{equation}
  C={R_x+\kappa R_t-\kappa^2D_\kappa R_t\over z^2}.
  \label{eq:C_recon}
\end{equation}
The metric functions are then
\begin{equation}
  f={1\over A},
  \qquad
  v={B\over A},
  \qquad
  h={1\over C+B^2/A}.
  \label{eq:fvh_recon}
\end{equation}
 \paragraph{}
\addtocounter{paragraph}{-1}

\paragraph{Null directions of the stationary block.}
\phantomsection\label{subsec:causal_readout_method}

The reconstructed metric determines the null directions tangent
to the constant-\(z\) stationary block.  Writing
\(\lambda=\dd x/\dd t\), the null condition is
\begin{equation}
  g_{tt}+2\lambda g_{tx}+\lambda^2 g_{xx}=0 .
  \label{eq:projected_null_quadratic_method}
\end{equation}
For the metric in Eq.~\eqref{eq:main_metric_convention}, this is
\begin{equation}
  -f(z)+h(z)\bigl(\lambda+v(z)\bigr)^2=0 .
  \label{eq:null_condition_fhv_method}
\end{equation}
Hence the two projected null directions are
\begin{equation}
  \lambda_\pm(z)=-v(z)\pm\sqrt{{f(z)\over h(z)}} .
  \label{eq:lambda_pm_fhv_method}
\end{equation}
In terms of the reconstructed inverse block variables,
\begin{equation}
  A={1\over f},
  \qquad
  B={v\over f},
  \qquad
  C={1\over h}-{v^2\over f},
\end{equation}
this becomes
\begin{equation}
  \lambda_\pm(z)
  ={ -B(z)\pm\sqrt{B(z)^2+A(z)C(z)}\over A(z)} .
  \label{eq:lambda_pm_ABC_method}
\end{equation}
Thus the projected light cone is fixed algebraically once \(A,B,C\) have
been reconstructed.  Conversely,
\begin{equation}
  v(z)=-{\lambda_+(z)+\lambda_-(z)\over2},
  \qquad
  {f(z)\over h(z)}={\bigl(\lambda_+(z)-\lambda_-(z)\bigr)^2\over4}.
  \label{eq:shift_opening_from_roots_method}
\end{equation}
The midpoint of the two roots is the local stationary shift, while their
separation is the opening of the projected cone.

If the reconstructed patch contains a stationary horizon at \(z=z_h\),
with \(f(z_h)=0\) and finite \(h(z_h)\), the two projected null
directions coincide:
\begin{equation}
  \lambda_+(z_h)=\lambda_-(z_h)=-v(z_h)
\end{equation}

The common direction is tangent to the horizon generator
\(\partial_t-v(z_h)\partial_x\).

\subsection{Consistency and overlap conditions}\label{sec:consistency}

The boundary input is $L=4G_N S_{\ren}$ on an open interval domain belonging to one smooth HRT family.  The conditions below test whether the three Abel kernels define one stationary homogeneous metric block. They require a shared geodesic Hamiltonian, the correct dependence on the charge ratio at fixed radius, one common radial coordinate, and Lorentzian signature.  The asymptotically AdS normalization has already been imposed in Eqs.~\eqref{eq:one_over_z_ode}--\eqref{eq:fvh_recon}.

\paragraph{Shared Hamiltonian relation.}
Because the three Abel kernels arise from the same radial variable,
their definitions imply
\begin{align}
  X_\kappa-u^2F_\kappa-\kappa T_\kappa=0 .
\end{align}
Equivalently, after dividing by \(F_\kappa\),
\begin{align}
  R_x-\kappa R_t-u^2=0 .
\end{align}
This condition tests whether the independently inverted length and
endpoint kernels can originate from one geodesic Hamiltonian.

\paragraph{Dependence on the charge ratio at fixed radius.}
At a fixed value of $z$, Eqs.~\eqref{eq:Rt_identity} and \eqref{eq:Rx_identity} are linear in $\kappa$. A stationary homogeneous block therefore requires
\begin{equation}
  D_\kappa^2R_t=0,
  \qquad
  D_\kappa^2R_x=0.
  \label{eq:fixed_radius_linearity}
\end{equation}
A violation of either relation excludes the stationary homogeneous three-dimensional ansatz on that data domain.  It may be caused by explicit time or spatial dependence, an unresolved change of extremal surface, or a higher-dimensional transverse density that has not been separated.

\paragraph{Identification of a common radial coordinate.}
Define
\begin{equation}
  Q(\kappa,u)
  \equiv
  {F_\kappa(u)\over\sqrt{D_\kappa R_t(\kappa,u)}} .
\end{equation}
Using \(F_\kappa=\sqrt{A}/(zu_\kappa')\) and
\(D_\kappa R_t=z^2A\), a stationary metric requires
\begin{equation}
  \partial_u\left({1\over z}\right)=-Q .
\end{equation}
A single radial coordinate shared by all fixed-\(\kappa\) families
must also satisfy
\begin{equation}
  D_\kappa\left({1\over z}\right)=0 .
\end{equation}
Applying $D_\kappa$ to $\partial_u(1/z)=-Q$ and using $D_\kappa(1/z)=0$ gives
\begin{equation}
  -D_\kappa Q
  =[D_\kappa,\partial_u]\left({1\over z}\right).
\end{equation}
Since $D_\kappa=\partial_\kappa-(R_t/u)\partial_u$,
\begin{equation}
  [D_\kappa,\partial_u]
  =\partial_u\left({R_t\over u}\right)\partial_u,
\end{equation}
and therefore
\begin{equation}
  D_\kappa\log Q
  =\partial_u\left({R_t\over u}\right).
  \label{eq:common_radial_condition}
\end{equation}
This condition is required for the relations $z=z_\kappa(u)$ obtained at different values of $\kappa$ to refer to the same bulk radial coordinate.

\paragraph{Lorentzian signature.}
On the exterior radial patch considered here, the metric in
Eq.~\eqref{eq:main_metric_convention} must satisfy \(f>0\) and
\(h>0\).  Since
\begin{align}
  A={D_\kappa R_t\over z^2},
  \qquad
  B=-{D_\kappa R_x\over z^2},
  \qquad
  C={R_x+\kappa R_t-\kappa^2D_\kappa R_t\over z^2},
\end{align}
the condition \(f=1/A>0\) requires
\begin{align}
  D_\kappa R_t>0 .
\end{align}
Furthermore,
\begin{align}
  h^{-1}
  =C+{B^2\over A}
  ={1\over z^2D_\kappa R_t}
  \left[
  D_\kappa R_t
  \bigl(R_x+\kappa R_t-\kappa^2D_\kappa R_t\bigr)
  +(D_\kappa R_x)^2
  \right],
\end{align}
so \(h>0\) requires
\begin{align}
  D_\kappa R_t
  \bigl(R_x+\kappa R_t-\kappa^2D_\kappa R_t\bigr)
  +(D_\kappa R_x)^2>0 .
\end{align}

Collecting the local relations gives
\begin{equation}
\begin{gathered}
  R_x-\kappa R_t-u^2=0,
  \qquad
  D_\kappa R_x+R_t-\kappa D_\kappa R_t=0,\\
  D_\kappa^2R_t=0,
  \qquad
  D_\kappa^2R_x=0,
  \qquad
  D_\kappa\log Q=\partial_u(R_t/u),\\
  D_\kappa R_t>0,
  \qquad
  D_\kappa R_t(R_x+\kappa R_t-\kappa^2D_\kappa R_t)+(D_\kappa R_x)^2>0.
\end{gathered}
\label{eq:consistency_summary}
\end{equation}
These are the local consistency conditions imposed before the reconstructed functions are accepted as one stationary metric block.

\paragraph{Agreement between overlapping local reconstructions.}\label{sec:matching}
Because the inverse formula is local, the same smooth entropy function may be inverted on more than one range of interval sizes.  This does not introduce a second HRT saddle.  For example, fix a common open range of \(\kappa\).  The data with \(0<u_*<u_1\) reconstruct metric functions \((A_1,B_1,C_1)\) on a radial interval \(I_1\).  Repeating the calculation with the same entropy function but with the larger range \(0<u_*<u_2\), where \(u_2>u_1\), reconstructs \((A_2,B_2,C_2)\) on a deeper interval \(I_2\).  Since \(u_*=u_\kappa(z_*)\) is monotone on the selected family, the first set of turning points is contained in the second, and hence \(I_1\subset I_2\).  After the same radial gauge and asymptotic normalization have been imposed, the two calculations must agree on the radial interval reconstructed by both:
\begin{equation}
  A_1(z)=A_2(z),
  \qquad
  B_1(z)=B_2(z),
  \qquad
  C_1(z)=C_2(z),
  \qquad z\in I_1\cap I_2.
  \label{eq:overlap_matching}
\end{equation}
This provides a consistency check that the reconstructed metric does not depend on the particular local interval domain used in the inversion.

\subsection{Higher-dimensional strips}\label{subsec:higher_dimensional_strips}

 In higher dimensions, a homogeneous strip carries a transverse
area density.  For a strip finite in the \(x\)-direction and extended
along \(y_a\), write the equal-time spatial metric as
\begin{align}
  ds^2_{\rm spatial}
  =
  {1\over z^2}
  \left[
  G(z)dz^2
  +H_x(z)dx^2
  +\sum_{a=2}^{d-1}H_a(z)dy_a^2
  \right].
\end{align}
After dividing by the coordinate volume of the transverse
directions, its area functional is
\begin{equation}
  {\cal A}_x
  =
  \int dx\,
  W_x(z)
  \sqrt{H_x(z)+G(z)z'(x)^2},
  \qquad
  W_x(z)
  \equiv
  z^{-(d-1)}
  \prod_{a=2}^{d-1}\sqrt{H_a(z)} .
  \label{eq:static_strip_area_density}
\end{equation}
This is the length functional associated with the effective
two-dimensional metric
\begin{align}
  d\bar s_{x,\rm spatial}^2
  =
  W_x(z)^2
  \left[
  G(z)dz^2+H_x(z)dx^2
  \right].
\end{align}
A single \(x\)-oriented strip reconstructs only the dressed
combinations \(W_x^2G\) and \(W_x^2H_x\).  It cannot separate the radial
and longitudinal metric functions from the transverse area density
\(W_x\). In an isotropic spatial sector,
\(H_x=H_a\equiv H\), so the transverse density
\(W_x=z^{-(d-1)}H^{(d-2)/2}\) introduces no additional independent
function.  After fixing the radial gauge, the strip data can therefore
determine the remaining radial functions \(G\) and \(H\).  In an anisotropic
sector, independent strip orientations are required to separate the
transverse density.

The same dressing occurs for covariant strips in stationary
geometries.  The reduced metric for a strip finite along \(x\) is
\begin{align}
  d\bar s_x^2
  =
  W_x(z)^2
  \left[
  G(z)dz^2
  -
  N(z)^2dt^2
  +
  H_x(z)(dx+v_x(z)dt)^2
  \right].
\end{align}
If \(A_x,B_x,C_x\) denote the inverse-block functions before this
multiplication, the three-dimensional inversion reconstructs
\begin{equation}
  \bar A_x\equiv{A_x\over W_x^2},
  \qquad
  \bar B_x\equiv{B_x\over W_x^2},
  \qquad
  \bar C_x\equiv{C_x\over W_x^2}.
  \label{eq:barred_ABC_definition}
\end{equation}
The transverse factor cancels from the stationary shift,
\begin{equation}
  v_x
  =
  {\bar B_x\over\bar A_x}
  =
  {B_x\over A_x},
  \label{eq:shift_survives_density_dressing}
\end{equation}
whereas the lapse and spatial warp factors remain dressed until
\(W_x\) is fixed by symmetry or separated using additional region
data.  With \(n=D-2\) boundary spatial directions, fixing one common
radial scale leaves \(n-1\) relative anisotropy functions.  The
corresponding data count is summarized in
Table~\ref{tab:scope_inversion}.

\section{Examples}\label{sec:examples}

\subsection{Pure AdS, rotating BTZ, and a static warped geometry}\label{subsec:local_sufficiency}

\phantomsection\label{sec:worked_example}

The examples are organized by the number of radial functions that must be recovered.  Pure AdS fixes the Abel and radial normalizations.  Rotating BTZ tests the complete three-function stationary reconstruction.  A perturbative static warped geometry then shows explicitly how nonzero-time intervals determine the lapse and the spatial warp factor independently when the shift vanishes.

\medskip
\phantomsection\label{subsec:example_pure_ads}
\paragraph{Pure AdS.}

For the vacuum CFT on the line, the renormalized geodesic length is
\begin{equation}
  L(\Delta t,\Delta x)
  =
  2\log\left({\sqrt{\Delta x^2-\Delta t^2}\over \epsilon}\right),
  \qquad
  \Delta x^2>\Delta t^2 ,
  \label{eq:ads_length}
\end{equation}
up to an additive constant.  Its endpoint derivatives are
\begin{equation}
  E=-\partial_{\Delta t}L
  ={2\Delta t\over \Delta x^2-\Delta t^2},
  \qquad
  J=\partial_{\Delta x}L
  ={2\Delta x\over \Delta x^2-\Delta t^2}.
  \label{eq:ads_charges}
\end{equation}
Thus
\begin{equation}
  \kappa={E\over J}={\Delta t\over \Delta x},
  \qquad
  u_*={1\over J}
  ={ \Delta x^2-\Delta t^2\over 2\Delta x}
  ={\Delta x\over2}(1-\kappa^2).
  \label{eq:ads_kappa_u}
\end{equation}
At fixed \(\kappa\), the boundary input becomes
\begin{equation}
  \Delta x_\kappa(u_*)={2u_*\over1-\kappa^2},
  \qquad
  \Delta t_\kappa(u_*)={2\kappa u_*\over1-\kappa^2},
  \qquad
  L_\kappa(u_*)=2\log\left({2u_*\over
  \epsilon\sqrt{1-\kappa^2}}\right).
  \label{eq:ads_interval_fixed_kappa_v75}
\end{equation}
For comparison with the reconstructed result, the known AdS turning-point condition uses \(\Psi_\kappa=1-\kappa^2\):
\begin{equation}
  1-J^2z_*^2(1-\kappa^2)=0
  \quad\Longrightarrow\quad
  z_*={1\over J\sqrt{1-\kappa^2}}
  ={1\over2}\sqrt{\Delta x^2-\Delta t^2}
  ={u_*\over\sqrt{1-\kappa^2}}.
  \label{eq:ads_turning_point}
\end{equation}
The inverse calculation below recovers this relation from the boundary functions in Eq.~\eqref{eq:ads_interval_fixed_kappa_v75}.

The endpoint kernels are obtained by applying the inverse Abel formula directly to the input data:
\begin{align}
  T_\kappa(u)
  &={1\over\pi}{\dd\over\dd u}
  \int_0^u{y\,\Delta t_\kappa(y)\over\sqrt{u^2-y^2}}\,\dd y,
  \label{eq:ads_T_inverse}\\
  X_\kappa(u)
  &={1\over\pi}{\dd\over\dd u}
  \int_0^u{y\,\Delta x_\kappa(y)\over\sqrt{u^2-y^2}}\,\dd y.
  \label{eq:ads_X_inverse}
\end{align}
Using
\begin{equation}
  \int_0^u{y^2\,\dd y\over\sqrt{u^2-y^2}}={\pi u^2\over4},
\end{equation}
one obtains
\begin{equation}
  T_\kappa(u)={\kappa u\over1-\kappa^2},
  \qquad
  X_\kappa(u)={u\over1-\kappa^2}.
  \label{eq:ads_TX_kernels}
\end{equation}
Since \(2y\mathcal L_\kappa(y)=L_\kappa(y)\), the length kernel is likewise
\begin{equation}
  F_\kappa(u)
  ={1\over\pi}{\dd\over\dd u}
  \int_0^u{L_\kappa(y)\over\sqrt{u^2-y^2}}\,\dd y.
  \label{eq:ads_F_inverse}
\end{equation}
The logarithmic input satisfies
\begin{equation}
  \int_0^u{L_\kappa(y)\over\sqrt{u^2-y^2}}\,\dd y
  =\pi\log\left({u\over\epsilon\sqrt{1-\kappa^2}}\right),
\end{equation}
and hence
\begin{equation}
  F_\kappa(u)={1\over u}.
  \label{eq:ads_F_kernel}
\end{equation}
Therefore
\begin{equation}
  R_t={T_\kappa\over F_\kappa}
  ={\kappa u^2\over1-\kappa^2},
  \qquad
  R_x={X_\kappa\over F_\kappa}
  ={u^2\over1-\kappa^2}.
  \label{eq:ads_Rt_Rx}
\end{equation}
The characteristic derivative gives
\begin{equation}
  D_\kappa R_t={u^2\over1-\kappa^2},
  \qquad
  D_\kappa R_x=0.
  \label{eq:ads_directional_derivatives}
\end{equation}
For this input, the radial equation becomes
\begin{equation}
  \partial_u\left({1\over z}\right)
  =-{F_\kappa\over\sqrt{D_\kappa R_t}}
  =-{\sqrt{1-\kappa^2}\over u^2}.
\end{equation}
Thus the boundary-normalized solution is
\begin{equation}
  z={u\over\sqrt{1-\kappa^2}}.
  \label{eq:ads_z_of_u}
\end{equation}
Substitution into the reconstruction formula gives
\begin{equation}
  A=1,
  \qquad
  B=0,
  \qquad
  C=1,
\end{equation}
and therefore
\begin{equation}
  f=1,
  \qquad
  h=1,
  \qquad
  v=0.
  \label{eq:ads_metric_reconstructed}
\end{equation}
Thus the input interval functions determine the three kernels, reproduce the turning-point relation in Eq.~\eqref{eq:ads_turning_point}, and recover the Poincar\'e AdS metric.

\medskip
\phantomsection\label{subsec:example_btz}
\paragraph{Rotating BTZ.}

For rotating BTZ, introduce the chiral interval variables
\begin{equation}
  \Delta x^+\equiv\Delta x+\Delta t,
  \qquad
  \Delta x^-\equiv\Delta x-\Delta t .
\end{equation}
In units where the outer horizon is at \(z=1\), define
\begin{equation}
  \alpha_+\equiv{1+\Omega\over2},
  \qquad
  \alpha_-\equiv{1-\Omega\over2}.
\end{equation}
The renormalized interval length is
\begin{equation}
  L(\Delta t,\Delta x)
  =
  \log\left[
  {1\over\epsilon^2\alpha_+\alpha_-}
  \sinh(\alpha_+\Delta x^+)
  \sinh(\alpha_-\Delta x^-)
  \right],
  \label{eq:v49_BTZ_CFT_entropy}
\end{equation}
up to an additive normalization.  Define
\begin{equation}
  a_+\equiv\alpha_+\coth(\alpha_+\Delta x^+),
  \qquad
  a_-\equiv\alpha_-\coth(\alpha_-\Delta x^-).
\end{equation}
Then
\begin{equation}
  E=-\partial_{\Delta t}L=a_- -a_+,
  \qquad
  J=\partial_{\Delta x}L=a_+ +a_-.
  \label{eq:v49_EJ_BTZ_CFT}
\end{equation}
At fixed \(\kappa=E/J\) and \(u_*=1/J\),
\begin{equation}
  a_+={1-\kappa\over2u_*},
  \qquad
  a_-={1+\kappa\over2u_*}.
\end{equation}
Solving for the interval separations gives the input functions
\begin{align}
  \Delta x^+_\kappa(u_*)
  &={1\over\alpha_+}
  \operatorname{arctanh}\left({2\alpha_+u_*\over1-\kappa}\right),
  \label{eq:btz_xplus_fixed_kappa_v75}\\
  \Delta x^-_\kappa(u_*)
  &={1\over\alpha_-}
  \operatorname{arctanh}\left({2\alpha_-u_*\over1+\kappa}\right).
  \label{eq:btz_xminus_fixed_kappa_v75}
\end{align}
It is convenient to define
\begin{equation}
  d_\kappa\equiv1-\kappa^2,
  \qquad
  p_\kappa\equiv1+\Omega^2+2\kappa\Omega,
  \qquad
  \lambda_+\equiv{2\alpha_+\over1-\kappa},
  \qquad
  \lambda_-\equiv{2\alpha_-\over1+\kappa},
  \qquad
  s_\pm(u)\equiv\sqrt{1-\lambda_\pm^2u^2}.
  \label{eq:btz_lambda_s_def}
\end{equation}
Using \(\sinh(\operatorname{arctanh}x)=x/\sqrt{1-x^2}\), the fixed-\(\kappa\) length input becomes
\begin{equation}
  L_\kappa(u_*)
  =2\log\left({2u_*\over\epsilon\sqrt{d_\kappa}}\right)
  -{1\over2}\log\bigl(1-\lambda_+^2u_*^2\bigr)
  -{1\over2}\log\bigl(1-\lambda_-^2u_*^2\bigr).
  \label{eq:btz_fixed_kappa_length}
\end{equation}

The chiral kernels are defined by the inverse transform of the chiral endpoint data,
\begin{equation}
  K_\pm(u)
  \equiv{1\over\pi}{\dd\over\dd u}
  \int_0^u{y\,\Delta x^\pm_\kappa(y)\over\sqrt{u^2-y^2}}\,\dd y.
  \label{eq:btz_chiral_kernel_inverse}
\end{equation}
The identity
\begin{equation}
  {1\over\pi}{\dd\over\dd u}
  \int_0^u{y\,\operatorname{arctanh}(\lambda y)
  \over\sqrt{u^2-y^2}}\,\dd y
  ={\lambda u\over2\sqrt{1-\lambda^2u^2}}
\end{equation}
then gives
\begin{align}
  K_+(u)
  &={u\over(1-\kappa)s_+(u)},
  \label{eq:btz_Kplus_v75}\\
  K_-(u)
  &={u\over(1+\kappa)s_-(u)}.
  \label{eq:btz_Kminus_v75}
\end{align}
The time and space kernels follow from the chiral combinations,
\begin{equation}
  T_\kappa={K_+-K_-\over2},
  \qquad
  X_\kappa={K_++K_-\over2}.
  \label{eq:btz_TX_kernels_v75}
\end{equation}

The length kernel is obtained independently from the length input,
\begin{equation}
  F_\kappa(u)
  ={1\over\pi}{\dd\over\dd u}
  \int_0^u{L_\kappa(y)\over\sqrt{u^2-y^2}}\,\dd y.
  \label{eq:btz_F_kernel_inverse}
\end{equation}
Besides the vacuum logarithm used in the AdS example, one needs
\begin{equation}
  {1\over\pi}{\dd\over\dd u}
  \int_0^u{-\frac{1}{2}\log(1-\lambda^2y^2)
  \over\sqrt{u^2-y^2}}\,\dd y
  ={1\over2u}\left({1\over\sqrt{1-\lambda^2u^2}}-1\right).
\end{equation}
Substitution of Eq.~\eqref{eq:btz_fixed_kappa_length} gives
\begin{equation}
  F_\kappa(u)
  ={1\over2u}\left({1\over s_+(u)}+{1\over s_-(u)}\right).
  \label{eq:btz_F_kernel_v75}
\end{equation}
The independently reconstructed kernels satisfy
\begin{equation}
  X_\kappa-u^2F_\kappa-\kappa T_\kappa=0.
\end{equation}
Thus the kernels obtained separately from the length and endpoint data are
consistent with one geodesic Hamiltonian.  Their roles are nevertheless
different.  The chiral kernels determine the endpoint kernels
\(T_\kappa\) and \(X_\kappa\), whereas the length kernel \(F_\kappa\)
is needed to convert the Abel variable \(u\) into the bulk radial coordinate.

The ratios entering the metric reconstruction are
\begin{align}
  R_t
  &={u^2\big[(1+\kappa)s_--(1-\kappa)s_+\big]
  \over d_\kappa(s_++s_-)},
  \\
  R_x
  &={u^2\big[(1+\kappa)s_-+(1-\kappa)s_+\big]
  \over d_\kappa(s_++s_-)}.
  \label{eq:btz_Rt_Rx_explicit}
\end{align}
They obey \(R_x-\kappa R_t=u^2\) identically.  The endpoint kernels determine \(R_t\) and \(R_x\), whereas the
independently reconstructed length kernel \(F_\kappa\),
Eq.~\eqref{eq:btz_F_kernel_v75}, fixes the relation between the Abel
variable \(u\) and the radial coordinate through
\begin{equation}
  \partial_u\left({1\over z}\right)
  =
  -{F_\kappa\over\sqrt{D_\kappa R_t}}=
  -\frac{
  d_\kappa\,[s_+(u)+s_-(u)]^2
  }{
  2\sqrt{2}\,u^2s_+(u)s_-(u)
  \sqrt{
  d_\kappa+(1+\Omega^2)u^2
  +d_\kappa s_+(u)s_-(u)}
  }.
  \label{eq:btz_radial_integrand_from_kernels_v80}
\end{equation}
Integrating this equation at fixed \(\kappa\) gives
\begin{equation}
  {1\over z}
  =
  {\sqrt{
  d_\kappa+(1+\Omega^2)u^2
  +d_\kappa s_+(u)s_-(u)}
  \over\sqrt{2}\,u}
  +C_\kappa ,
  \label{eq:btz_inverse_radius_integrated_v80}
\end{equation}
where \(C_\kappa\) is independent of \(u\).

Near the asymptotic boundary,
\(A,C\to1\) and \(B\to0\), so that
\(\Psi_\kappa\to1-\kappa^2\equiv d_\kappa\).
Since \(u=z\sqrt{\Psi_\kappa}\), the required asymptotic behavior is
\[
  {1\over z}\sim{\sqrt{d_\kappa}\over u}.
\]
The first term in Eq.~\eqref{eq:btz_inverse_radius_integrated_v80}
already has this behavior as \(u\to0\), fixing \(C_\kappa=0\).
Hence
\begin{equation}
  {1\over z}
  =
  {\sqrt{
  d_\kappa+(1+\Omega^2)u^2
  +d_\kappa s_+(u)s_-(u)}
  \over\sqrt{2}\,u}.
  \label{eq:btz_inverse_radius_from_kernels_v80}
\end{equation}

To eliminate the auxiliary square roots, their definitions give
\begin{align}
  d_\kappa^2s_+^2s_-^2
  &=
  d_\kappa^2
  \left[
  1-(\lambda_+^2+\lambda_-^2)u^2
  +\lambda_+^2\lambda_-^2u^4
  \right]
  \nonumber\\
  &=
  \left[d_\kappa+(1+\Omega^2)u^2\right]^2
  -4u^2\left[p_\kappa+\Omega^2u^2\right],
  \label{eq:btz_s_product_identity_v80}
\end{align}
where the second equality follows by inserting the explicit
\(\lambda_\pm\) in Eq.~\eqref{eq:btz_lambda_s_def}.
Equation~\eqref{eq:btz_inverse_radius_from_kernels_v80} also implies
\begin{equation}
  d_\kappa s_+s_-
  =
  {2u^2\over z^2}
  -d_\kappa-(1+\Omega^2)u^2 .
\end{equation}
Combining these two relations yields
\begin{align}
  &u^2
  -z^2\left[d_\kappa+(1+\Omega^2)u^2\right]
  +z^4\left[p_\kappa+\Omega^2u^2\right]
  =0
  \nonumber\\
  &\Longleftrightarrow\quad
  u^2(1-z^2)(1-\Omega^2z^2)
  =
  z^2\left[d_\kappa-p_\kappa z^2\right].
  \label{eq:btz_u_z_relation_from_kernels_v75}
\end{align}
The solution continuously connected to the asymptotic boundary is
\begin{equation}
  z^2
  =
  {d_\kappa+(1+\Omega^2)u^2
  -\sqrt{
  \left[d_\kappa+(1+\Omega^2)u^2\right]^2
  -4u^2\left[p_\kappa+\Omega^2u^2\right]}
  \over
  2\left[p_\kappa+\Omega^2u^2\right]} .
  \label{eq:btz_z_of_u_from_kernels_v75}
\end{equation}
At fixed reconstructed radius, the characteristic derivative gives
\begin{equation}
  D_\kappa R_t
  ={z^2\over(1-z^2)(1-\Omega^2z^2)},
  \qquad
  D_\kappa R_x
  =-{\Omega z^4\over(1-z^2)(1-\Omega^2z^2)}.
  \label{eq:btz_DRt_DRx_from_kernels_v75}
\end{equation}
Consequently,
\begin{align}
  A&={1\over(1-z^2)(1-\Omega^2z^2)},
  \\
  B&={\Omega z^2\over(1-z^2)(1-\Omega^2z^2)},
  \\
  C&=1-{\Omega^2z^4\over(1-z^2)(1-\Omega^2z^2)}.
\end{align}
The metric functions reconstructed from the boundary input are therefore
\begin{equation}
  f(z)=(1-z^2)(1-\Omega^2z^2),
  \qquad
  h(z)=1,
  \qquad
  v(z)=\Omega z^2.
  \label{eq:btz_final_from_kernels_v75}
\end{equation}
In particular,
\begin{equation}
  v(z)=-{D_\kappa R_x\over D_\kappa R_t}=\Omega z^2,
\end{equation}
so the left--right asymmetry of the input entropy fixes the complete radial shift profile, including its horizon value.

\paragraph{A static warped geometry.}

Pure AdS and rotating BTZ both have \(h(z)=1\) in the chosen gauge.  To test whether the inverse map can recover an independent spatial warp factor \(h(z)\neq1\), while keeping the stationary shift absent, consider the perturbative metric
\begin{equation}
  f(z)=1-\varepsilon\mu z^2+O(\varepsilon^2),
  \qquad
  h(z)=1-\varepsilon\nu z^2+O(\varepsilon^2),
  \qquad
  v(z)=0,
  \label{eq:static_warped_pert_metric}
\end{equation}
where \(\mu\) and \(\nu\) are independent and \(\varepsilon\) is a bookkeeping parameter.  Then
\begin{equation}
  A=1+\varepsilon\mu z^2,
  \qquad
  B=0,
  \qquad
  C=1+\varepsilon\nu z^2
  \qquad
  +O(\varepsilon^2).
\end{equation}
Let \(d_\kappa\equiv1-\kappa^2\).  Expanding the endpoint integrals generated by this geometry at fixed \(\kappa\) gives the boundary input
\begin{align}
  L_\kappa(u_*)
  &=2\log\left({2u_*\over\epsilon_{\rm UV}\sqrt{d_\kappa}}\right)
  +\varepsilon{\mu(1+\kappa^2)-2\nu\over d_\kappa^2}u_*^2
  +O(\varepsilon^2),
  \label{eq:static_warped_input_L}\\
  \Delta t_\kappa(u_*)
  &={2\kappa u_*\over d_\kappa}
  +{2\varepsilon\kappa\big[\mu(3+\kappa^2)-4\nu\big]
  \over3d_\kappa^3}u_*^3
  +O(\varepsilon^2),
  \label{eq:static_warped_input_t}\\
  \Delta x_\kappa(u_*)
  &={2u_*\over d_\kappa}
  +{2\varepsilon\big[\mu(1+3\kappa^2)-2\nu(1+\kappa^2)\big]
  \over3d_\kappa^3}u_*^3
  +O(\varepsilon^2).
  \label{eq:static_warped_input_x}
\end{align}
At \(\kappa=0\), the corrections depend only on \(\mu-2\nu\); equal-time data therefore cannot determine the two coefficients independently.

Applying the inverse formulas to Eqs.~\eqref{eq:static_warped_input_L}--\eqref{eq:static_warped_input_x},
\begin{align}
  F_\kappa(u)
  &={1\over\pi}{\dd\over\dd u}
  \int_0^u{L_\kappa(y)\over\sqrt{u^2-y^2}}\,\dd y,
  \\
  T_\kappa(u)
  &={1\over\pi}{\dd\over\dd u}
  \int_0^u{y\,\Delta t_\kappa(y)\over\sqrt{u^2-y^2}}\,\dd y,
  \\
  X_\kappa(u)
  &={1\over\pi}{\dd\over\dd u}
  \int_0^u{y\,\Delta x_\kappa(y)\over\sqrt{u^2-y^2}}\,\dd y,
\end{align}
and using
\begin{equation}
  {1\over\pi}{\dd\over\dd u}
  \int_0^u{y^2\,\dd y\over\sqrt{u^2-y^2}}={u\over2},
  \qquad
  {1\over\pi}{\dd\over\dd u}
  \int_0^u{y^4\,\dd y\over\sqrt{u^2-y^2}}={3u^3\over4},
\end{equation}
one obtains
\begin{align}
  F_\kappa(u)
  &={1\over u}
  +\varepsilon{\mu(1+\kappa^2)-2\nu\over2d_\kappa^2}u
  +O(\varepsilon^2),
  \label{eq:static_warped_F_kernel}\\
  T_\kappa(u)
  &={\kappa u\over d_\kappa}
  +\varepsilon{\kappa\big[\mu(3+\kappa^2)-4\nu\big]
  \over2d_\kappa^3}u^3
  +O(\varepsilon^2),
  \label{eq:static_warped_T_kernel}\\
  X_\kappa(u)
  &={u\over d_\kappa}
  +\varepsilon{\mu(1+3\kappa^2)-2\nu(1+\kappa^2)
  \over2d_\kappa^3}u^3
  +O(\varepsilon^2).
  \label{eq:static_warped_X_kernel}
\end{align}
Their ratios are
\begin{align}
  R_t
  &={\kappa u^2\over d_\kappa}
  \left[1+\varepsilon{\mu-\nu\over d_\kappa^2}u^2\right]
  +O(\varepsilon^2),
  \\
  R_x
  &={u^2\over d_\kappa}
  \left[1+\varepsilon{\kappa^2(\mu-\nu)\over d_\kappa^2}u^2\right]
  +O(\varepsilon^2).
  \label{eq:static_warped_ratios}
\end{align}
The characteristic derivative gives
\begin{equation}
  D_\kappa R_t
  ={u^2\over d_\kappa}
  \left[1+\varepsilon{\mu-\nu\over d_\kappa^2}u^2\right]
  +O(\varepsilon^2),
  \qquad
  D_\kappa R_x=0.
  \label{eq:static_warped_D_ratios}
\end{equation}
For this geometry, the radial equation becomes
\begin{equation}
  \partial_u\left({1\over z}\right)
  =-{F_\kappa\over\sqrt{D_\kappa R_t}}
  =-{\sqrt{d_\kappa}\over u^2}
  \left[1-\varepsilon{\nu-\mu\kappa^2\over2d_\kappa^2}u^2\right]
  +O(\varepsilon^2),
\end{equation}
and hence
\begin{equation}
  z={u\over\sqrt{d_\kappa}}
  \left[1-\varepsilon{\nu-\mu\kappa^2\over2d_\kappa^2}u^2\right]
  +O(\varepsilon^2).
  \label{eq:static_warped_z_u}
\end{equation}
Finally,
\begin{equation}
  A={D_\kappa R_t\over z^2}=1+\varepsilon\mu z^2,
  \qquad
  B=-{D_\kappa R_x\over z^2}=0,
  \qquad
  C={R_x\over z^2}=1+\varepsilon\nu z^2,
\end{equation}
so
\begin{equation}
  f(z)=1-\varepsilon\mu z^2,
  \qquad
  h(z)=1-\varepsilon\nu z^2,
  \qquad
  v(z)=0
  \qquad +O(\varepsilon^2).
  \label{eq:static_warped_reconstructed}
\end{equation}
The \(\kappa\)-dependence of the input determines \(\mu\) and \(\nu\) independently; this is the information absent from the equal-time entanglement entropy input.

\subsection{An Einstein--scalar black brane}\label{subsec:example_scalar}

\phantomsection\label{sec:rotating_scalar_exact}

This example has two purposes.  First, it checks the inverse map on a non-BTZ metric with two nontrivial static functions.  Second, after the metric has been reconstructed, it shows what additional information follows only after a minimal Einstein--scalar action is imposed.

\medskip
\paragraph{Exact branch in the reconstruction gauge.}
Consider the three-dimensional Einstein--scalar action
\begin{equation}
  I={1\over16\pi G}\int d^3x\sqrt{-g}\left[
  R-{1\over2}(\partial\phi)^2-V(\phi)
  \right] .
  \label{eq:einstein_scalar_action_v52}
\end{equation}
For
\begin{equation}
  V(\phi)=
  -{2\left(3+9s+13s^2-s^3\right)\over3(1+s)^2},
  \qquad
  s\equiv\tan^2{\phi\over2},
  \label{eq:exact_scalar_potential_v52}
\end{equation}
there is an exact static solution
\begin{align}
  ds^2&={dr^2\over F(r)}+e^{2A(r)}\left[-F(r)d\tilde t^2+d\tilde x^2\right],
  \\
  A(r)&=r-{1\over2}\log\left(1+e^{-2r}\right),
  \\
  F(r)&=1-{2\over3}e^{-2r}-{1\over3}e^{-4r},
  \\
  \phi(r)&=2\arctan e^{-r}.
  \label{eq:exact_static_scalar_solution}
\end{align}
The boundary is at \(r\to\infty\) and the horizon is at \(r=0\).  To put the metric in the radial gauge of Eq.~\eqref{eq:main_metric_convention}, choose \(z(r)\) to satisfy
\begin{equation}
  {\dd\over\dd r}\left({1\over z}\right)=e^{A(r)},
  \qquad
  z\sim e^{-r}\quad(r\to\infty).
  \label{eq:scalar_radial_gauge_map}
\end{equation}
For the exact function \(A(r)\), this integrates to
\begin{equation}
  {1\over z}=\sqrt{1+e^{2r}},
  \qquad
  e^{-2r}={z^2\over1-z^2}.
  \label{eq:scalar_r_z_exact}
\end{equation}
The metric and scalar therefore become
\begin{equation}
  ds^2={1\over z^2}\left[-f(z)d\tilde t^2+{dz^2\over f(z)}+h(z)d\tilde x^2\right],
  \label{eq:scalar_metric_main_gauge}
\end{equation}
with
\begin{equation}
  f(z)=(1-2z^2)\left(1-{2\over3}z^2\right),
  \qquad
  h(z)=(1-z^2)^2,
  \qquad
  \phi(z)=2\arcsin z.
  \label{eq:scalar_f_h_phi_exact}
\end{equation}
The horizon is at \(z_h=1/\sqrt2\).  Near the boundary,
\begin{equation}
  A_{\rm inv}(z)\equiv{1\over f(z)}=1+{8\over3}z^2+O(z^4),
  \qquad
  C_{\rm inv}(z)\equiv{1\over h(z)}=1+2z^2+O(z^4).
  \label{eq:scalar_inverse_near_boundary}
\end{equation}

\paragraph{Kernel inversion and recovery of the metric.}
Let \(L_\kappa(u_*)\), \(\Delta\tilde t_\kappa(u_*)\), and \(\Delta\tilde x_\kappa(u_*)\) denote the exact boundary data of this branch.  The kernels are defined from these input functions by
\begin{align}
  F_\kappa(u)
  &\equiv{1\over\pi}{\dd\over\dd u}
  \int_0^u{L_\kappa(y)\over\sqrt{u^2-y^2}}\,\dd y,
  \\
  T_\kappa(u)
  &\equiv{1\over\pi}{\dd\over\dd u}
  \int_0^u{y\,\Delta\tilde t_\kappa(y)\over\sqrt{u^2-y^2}}\,\dd y,
  \\
  X_\kappa(u)
  &\equiv{1\over\pi}{\dd\over\dd u}
  \int_0^u{y\,\Delta\tilde x_\kappa(y)\over\sqrt{u^2-y^2}}\,\dd y.
  \label{eq:scalar_kernel_inverse_definitions}
\end{align}
Near the boundary, the exact scalar solution has
\(A_{\rm inv}=1+\frac{8}{3} z^2+O(z^4)\) and
\(C_{\rm inv}=1+2z^2+O(z^4)\).
It therefore matches the general static warped expansion with
\(\mu=8/3\) and \(\nu=2\) at the first nontrivial order. The first nontrivial terms of the exact boundary input are obtained from Eqs.~\eqref{eq:static_warped_input_L}--\eqref{eq:static_warped_input_x} by setting \(\mu=8/3\) and \(\nu=2\):
\begin{align}
  L_\kappa(u_*)
  &=2\log\left({2u_*\over\epsilon_{\rm UV}\sqrt{d_\kappa}}\right)
  +{4(2\kappa^2-1)\over3d_\kappa^2}u_*^2+O(u_*^4),
  \\
  \Delta\tilde t_\kappa(u_*)
  &={2\kappa u_*\over d_\kappa}
  +{16\kappa^3\over9d_\kappa^3}u_*^3+O(u_*^5),
  \\
  \Delta\tilde x_\kappa(u_*)
  &={2u_*\over d_\kappa}
  +{8(3\kappa^2-1)\over9d_\kappa^3}u_*^3+O(u_*^5).
  \label{eq:scalar_boundary_input_expansion}
\end{align}
Applying Eq.~\eqref{eq:scalar_kernel_inverse_definitions} reproduces Eq.~\eqref{eq:scalar_inverse_near_boundary}.  For this static solution, the general characteristic variable $u_\kappa =z \sqrt{\Psi_\kappa}$ reduces to
\begin{equation}
  u_\kappa(z)
  = z\sqrt{{1\over h(z)}-{\kappa^2\over f(z)}}.
  \label{eq:scalar_exact_u_kappa}
\end{equation}
The inverse transforms of the exact endpoint data give
\begin{align}
  F_\kappa\bigl(u_\kappa(z)\bigr)
  &={1\over z\sqrt{f(z)}\,u_\kappa'(z)},
  \\
  T_\kappa\bigl(u_\kappa(z)\bigr)
  &={\kappa z\over f(z)^{3/2}u_\kappa'(z)},
  \\
  X_\kappa\bigl(u_\kappa(z)\bigr)
  &={z\over h(z)\sqrt{f(z)}\,u_\kappa'(z)}.
  \label{eq:scalar_exact_parametric_kernels}
\end{align}
Although the endpoint integrals are not elementary, Eq.~\eqref{eq:scalar_exact_parametric_kernels} is an exact evaluation of their Abel inverses.  Their ratios are
\begin{equation}
  R_t={\kappa z^2\over f(z)},
  \qquad
  R_x={z^2\over h(z)}.
  \label{eq:scalar_exact_ratios}
\end{equation}
The characteristic derivative gives
\begin{equation}
  D_\kappa R_t={z^2\over f(z)},
  \qquad
  D_\kappa R_x=0.
  \label{eq:scalar_exact_directional}
\end{equation}

The exact kernels satisfy
\begin{equation}
  {F_\kappa\over\sqrt{D_\kappa R_t}}
  ={1\over z^2u_\kappa'(z)}=-\partial_u\left(\frac{1}{z}\right),
\end{equation}
which is consistent with the radial reconstruction equation
\(\partial_u(1/z)=-F_\kappa/\sqrt{D_\kappa R_t}\).
Thus the exact parametric kernels satisfy the nonlinear inverse
relations throughout the branch. Finally, we have
\begin{equation}
  A={D_\kappa R_t\over z^2}={1\over f(z)},
  \qquad
  B=0,
  \qquad
  C={R_x\over z^2}={1\over h(z)},
\end{equation}
and Eq.~\eqref{eq:scalar_f_h_phi_exact} is recovered exactly. 

\paragraph{Conditional reconstruction of scalar data.}
\phantomsection\label{subsec:scalar_hkll_relation_v52}
\phantomsection\label{sec:scalar_potential_reconstruction_detail}

The Abel inversion reconstructs the metric without choosing the matter theory. We can also reconstruct the scalar potential by imposing the minimal action in Eq.~\eqref{eq:einstein_scalar_action_v52}.  From the reconstructed functions, one first returns to the rest-frame coordinate through
\begin{equation}
  e^{2A(r(z))}={h(z)\over z^2},
  \qquad
  F(r(z))={f(z)\over h(z)},
  \qquad
  {dr\over dz}=-{1\over z\sqrt{h(z)}}.
  \label{eq:scalar_rest_functions_from_metric}
\end{equation}
For the exact branch this gives Eq.~\eqref{eq:scalar_r_z_exact}.  The Einstein equations following from Eq.~\eqref{eq:einstein_scalar_action_v52} are
\begin{equation}
  R_{\mu\nu}={1\over2}\partial_\mu\phi\partial_\nu\phi+V(\phi)g_{\mu\nu},
  \label{eq:Einstein_scalar_Rmunu_v53}
\end{equation}
and reduce in the static ansatz to
\begin{align}
  \phi'(r)^2&=-2A''(r),
  \label{eq:phi_from_A_v53}\\
  F''(r)+2A'(r)F'(r)&=0,
  \label{eq:F_eq_A_v53}\\
  V(\phi(r))&=-\left[F(A''+2A'^2)+A'F'\right].
  \label{eq:V_from_AF_v53}
\end{align}
Thus the reconstructed metric determines \(|\phi'|\) and \(V\) along the solution branch.  The sign of \(\phi'\) is a branch choice; monotonicity and \(\phi\to0\) at the boundary select the branch used here.

For the exact solution, let
\begin{equation}
  y\equiv e^{-2r}=\tan^2{\phi\over2}.
\end{equation}
Then
\begin{equation}
  A'(r)={1+2y\over1+y},
  \qquad
  A''(r)=-{2y\over(1+y)^2},
\end{equation}
and
\begin{equation}
  \phi'(r)^2={4y\over(1+y)^2}=-2A''(r).
\end{equation}
Moreover,
\begin{equation}
  F'(r)={4\over3}y(1+y),
  \qquad
  F''(r)=-{8\over3}y(1+2y),
\end{equation}
so \(F''+2A'F'=0\).  Substituting these functions into
Eq.~\eqref{eq:V_from_AF_v53} gives
\begin{align}
V(r)
&=
-\left[
\left(1-\frac{2}{3}y-\frac{1}{3}y^2\right)
\frac{2+6y+8y^2}{(1+y)^2}
+\frac{4}{3}y(1+2y)
\right]
\nonumber\\
&=
-\frac{2(3+9y+13y^2-y^3)}{3(1+y)^2}.
\end{align}
Since \(y=\tan^2(\phi/2)\), the potential reconstructed on the field interval sampled by the branch reproduces Eq.~\eqref{eq:exact_scalar_potential_v52}.  Near the boundary, \(V(\phi)=-2-\phi^2/2+O(\phi^4)\), corresponding to the BF-saturating mass \(m^2=-1\).

The geometric inverse map therefore recovers the complete backreacted metric.  The scalar magnitude and the potential are reconstructed only conditionally, after the Einstein--scalar equations and a monotone scalar branch have been specified.

\section{Discussion}
\label{sec:discussion}

The main result is an extension of analytic holographic metric inversion from static equal-time data to covariant HRT data. The covariant interval entropy $S_{\ren}(\Delta t,\Delta x)$ supplies two Hamilton--Jacobi charges, $E$ and $J$.  Their ratio $\kappa=E/J$ labels extremal curves with different time--space orientations, while $u_*=1/J$ varies the turning point within each fixed-$\kappa$ family.  The resulting three Abel kernels determine the three independent components of the stationary inverse block and, after the common radial coordinate is fixed, reconstruct $f(z)$, $h(z)$, and $v(z)$.  Pure AdS fixes the normalization of the Abel and radial variables.  Rotating BTZ tests the nonzero shift profile $v(z)$. The static warped example tests an independent spatial warp factor $h(z)\neq1$, and the Einstein--scalar example first reconstructs the metric functions $f(z)$ and $h(z)$, which are deformed by the scalar field. Furthermore, the scalar field $\phi(z)$ and potential $V(\phi)$ are obtained only after imposing the Einstein--scalar field equations.

The result is local in interval space and in the bulk.  It assumes $J\neq0$, a nondegenerate map from the endpoint separations to $(\kappa,u_*)$, and a monotone characteristic variable on the reconstructed radial patch.  The formula does not yield complete reconstruction when there's a caustic or a degenerate turning point. When the same smooth entropy function is inverted on overlapping interval data, the reconstructed functions must agree on the shared radial range as in Eq.~\eqref{eq:overlap_matching}. If several extremal surfaces compete, the physical entropy selects the one with the least area.  The inversion applies where that selected extremum varies smoothly, but the physical entropy alone does not determine the other extremal surface across an HRT phase transition.

The principal structural limitation is the symmetry assumption that makes the HRT problem effectively radial.  Stationarity and homogeneity imply
\begin{equation}
  S(t_L,x_L;t_R,x_R)=S(\Delta t,\Delta x),
\end{equation}
whereas a time-dependent or spatially inhomogeneous state generally has
\begin{equation}
  S=S(T,X;\Delta t,\Delta x),
  \qquad
  T\equiv{t_L+t_R\over2},
  \qquad
  X\equiv{x_L+x_R\over2}.
  \label{eq:midpoint_entropy_dependence}
\end{equation}
The same distinction appears in the bulk.  For     $p_\mu=g_{\mu\nu}\dot X^\nu$,
\begin{equation}
  {d p_t\over ds}
  ={1\over2}\,\partial_t g_{\mu\nu}\dot x^\mu\dot x^\nu,
  \qquad
  {d p_x\over ds}
  ={1\over2}\,\partial_x g_{\mu\nu}\dot x^\mu\dot x^\nu .
  \label{eq:nonconserved_endpoint_momenta}
\end{equation}
Thus, $E=-p_t$ and $J=p_x$ cease to be conserved when the corresponding metric dependence is present, and the family can no longer be organized by the two constants $(\kappa,u_*)$. A time-dependent or inhomogeneous extension must retain the midpoint variables $T$ and $X$ in the inverse problem. Developing such a transform would test whether the present construction can be extended to a local reconstruction with explicit center-time or spatial dependence.

Higher dimensions introduce a different obstruction.  A strip reconstructs a radial block multiplied by the transverse area density of the extremal surface. Isotropy expresses that density in terms of the same warp factor that controls the finite strip direction, so no additional independent function remains after the radial gauge is fixed.  In an anisotropic geometry, one strip orientation cannot separately determine the radial metric functions and the transverse area density.  A concrete next step is to combine several strip orientations or region shapes and determine when their reconstructed blocks satisfy one common higher-dimensional metric.  The function count then identifies which additional entanglement observables are required.

In addition, the inverse map is kinematic and classical. It reconstructs a metric compatible with the HRT data but does not determine the gravitational action or the bulk matter content.  The Einstein--scalar example illustrates the additional assumptions required to infer a scalar profile and potential.  It would be useful to supplement the metric inversion with dynamical boundary information that constrains perturbations of the reconstructed geometry and the equations they obey. Higher-curvature entropy functionals, bulk entanglement corrections, quantum extremal surfaces, and finite-$N$ effects change the functional being inverted and require a separate extension of the present construction.

Finally, differentiation of the entropy followed by Abel inversion amplifies numerical uncertainty. Applying the formulas to numerical HRT data or to entropies obtained from a microscopic model will require regularized inversion, uncertainty propagation, and numerical tests of the consistency conditions in Sec.~\ref{sec:consistency} using discretely sampled entropy data. Such an implementation could distinguish uncertainty in the input from failure of the stationary metric ansatz. Within the symmetry class treated here, the inverse map therefore provides both an explicit reconstruction and falsifiable tests of whether proposed entropy data can arise from one stationary bulk metric.

\section*{Acknowledgments}
We especially thank Hun Jang, Hyun Seok Yang, Jae Hyuk Oh, Sang Jin Sin, and Keun Young Kim for their valuable comments and discussions.

\bibliographystyle{unsrtnat}
\bibliography{bibliography}

\end{document}